\documentclass{iopconfser}

\usepackage{amsmath}
\usepackage{amssymb}
\usepackage{braket}

\begin{document}

\title{Induced quantum gravity from QFT vector models}

\author{Matti Raasakka$^{1}$}

\affil{$^1$Department of Electrical Engineering, Aalto University, Espoo, Finland}

\email{matti.raasakka@aalto.fi}

\begin{abstract}
QFT vector models are a newly developed approach to quantum gravity, which are based on induced gravity in discrete spacetimes. Here we review some basic definitions and properties of these models and point out directions for future research.
\end{abstract}

\section{Introduction and motivation}
QFT vector models are a new approach to quantum gravity (QG), originally introduced in \cite{raasakka25}, which combine insights and techniques from various other current approaches, such as causal dynamical triangulations \cite{Ambjorn12}, random tensor models \cite{Gurau2024} and tensor networks \cite{Orus19,May17,Milsted18}. The basic idea is to first formulate quantum field theory (QFT) on simplicial manifolds via tensor network techniques. The gravitational action for the manifold then arises (if things go well) as an effective low-energy action from QFT amplitudes instead of being put in as a fundamental input --- so-called `induced gravity'. This is an old idea, already pointed out by Sakharov in the 60's \cite{Sakharov68,Visser02}, but it has not received very much attention, perhaps because of the infamous cosmological constant problem: The effective cosmological constant comes out around 100 orders of magnitude too large from the calculations! However, for the QFT vector model, we will see that reformulating the QFT model as a state-sum model will allow us to renormalize the vacuum energy via a coupling constant redefinition. After defining the general class of models, we will also discuss the emergence of gravitational amplitudes and the classical and continuum limits of the theory.

\section{Definition of QFT vector models}

\subsection{Kinematical boundary Hilbert space for a single simplex}
Let us consider a regular flat $d$-simplex $\Delta$ of proper edge time/length $l_\Delta$ equipped inside with a bulk Minkowski metric. Its boundary consists of $d+1$ $(d-1)$-simplices we call the \emph{faces} of the simplex, which can be either space- or timelike. (We do not consider null faces here, since we want them to have a non-zero invariant size.) The effect on quantum fields by a classical source $g_f \in L^2(f,\mathbb{C})$ on face $f$ is represented by the smeared field operator $\hat{\Phi}[g_f] = \int_f g_f \hat{\Phi}$, where $\hat{\Phi}$ is the usual point-wise field operator.  (We will suppress the notation for other degrees of freedom, such as spin or charge.) On each face $f$ we introduce a basis of functions $\zeta_k^{(f)} \in L^2(f,\mathbb{C})$, the \emph{boundary modes} of the field, where $k$ is a discrete index labeling the basis. Any classical source $g_f$ on $f$ can then be decomposed as a linear combination of the mode functions as $g_f = \sum_k \tilde{g}_k \zeta_k^{(f)}$,  and the same for $\hat{\Phi}[g_f] = \sum_k \tilde{g}_k \hat{\Phi}[\zeta_k^{(f)}]$ due to linearity. To consider superpositions of classical boundary sources, and thus obtain the boundary Hilbert space, we construct Fock spaces $\mathcal{H}_f$ for each face $f$ based on the field operators $\hat{\Phi}_k^{(f)} := \hat{\Phi}[\zeta_k^{(f)}]$ for the boundary field modes. The number states, which provide an orthonormal basis for the Fock space $\mathcal{H}_f$, can be constructed by acting on the vacuum state $\ket{\text{vac}}$ with the mode operators: For example, $|n_k,n_l\rangle = S\left( \hat{\Phi}_k^{(f)} \right)^{n_k} \left( \hat{\Phi}_l^{(f)} \right)^{n_l} \ket{\text{vac}}$, where $S$ is the appropriate (anti-)symmetrization operation implementing the statistics, represents a state with $n_k$ and $n_l$ excitations in the modes $k$ and $l$, respectively, on the face $f$. Now, the \emph{kinematical boundary Hilbert space} for the simplex is obtained simply as the tensor product $\mathcal{H}_\Delta^{(\text{kin})} = \otimes_f \mathcal{H}_f$. The term `kinematical' refers to the fact that this Hilbert space still contains unphysical states, which are not compatible with the dynamics of the model. We denote the kinematical boundary basis states as $\otimes_f |{\bf n}_f\rangle = |{\bf n}_1 \ldots {\bf n}_{d+1}\rangle$, where ${\bf n}_f = (n_k^{(f)})_k$ denotes the vector of excitation numbers $n_k^{(f)}$ for the face $f$.

\subsection{Single simplex amplitudes and the amplitude tensor}
Following the general boundary formulation of QFT \cite{Oeckl08,Oeckl24}, we now associate probability amplitudes to boundary states, giving rise to the amplitude map $A: \mathcal{H}_\Delta^{(\text{kin})} \rightarrow \mathbb{C}$. As the amplitude map is linear, its values on the Fock basis states can be interpreted as constituting a tensor $A_{{\bf n}_1 \ldots {\bf n}_{d+1}} \equiv A(|{\bf n}_1 \ldots {\bf n}_{d+1}\rangle)$ with the index ${\bf n}_f$ labeling a Fock basis state for the face $f$ of the simplex. The \emph{physical boundary Hilbert space} $\mathcal{H}_\Delta^{(\text{phys})}$ is then obtained by quotienting out the unphysical boundary states with zero amplitude  and completing in the Hilbert space norm, $\mathcal{H}_\Delta^{(\text{phys})} = \overline{\mathcal{H}_\Delta^{(\text{kin})}/\text{ker}(A)}$.

For the explicit construction of the amplitudes we assume the simplex to be embedded into a flat Minkowski spacetime. We can then express the amplitudes as Minkowski vacuum expectation values, and thus in terms of the Minkowski spacetime $n$-point function $G_n$ as
\begin{align}\label{eq:amptens}
    A_{{\bf n}_1 \ldots {\bf n}_{d+1}} = \langle\text{vac}| TS\left( \prod_{f,k} \left( \hat{\Phi}_k^{(f)} \right)^{n_k^{(f)}} \right) |\text{vac}\rangle  = \left[ \prod_f \prod_k \int_f \text{d}x_{f,k} \zeta_k^{(f)}(x_{f,k}) \right] G_n(x_{f,k}) \,,
\end{align}
where $T$ denotes the time-ordering, and $n = \sum_{f,k} n_k^{(f)}$ is the total number of boundary excitations. The choice of Minkowski vacuum amplitudes can be motivated by the equivalence principle: The local physics inside a single simplex should look like that in flat spacetime.\footnote{Even though physically motivated, this choice can be contested, because we are essentially ignoring the boundedness of the simplex. Work to consider other possible options, such as developed in \cite{Oeckl12}, is currently under way.}

\subsection{Gluing of simplicial amplitudes}
We can now start gluing simplices together to form simplicial manifolds and the associated boundary amplitudes. To glue two simplices together along a common face, we introduce the gluing tensor $G_{{\bf n}{\bf n}'}$, which takes care of matching the orientations of the two faces when we identify the two Fock bases ${\bf n},{\bf n}'$.\footnote{In general, we will have several different amplitude and gluing tensors corresponding to different spatiotemporal orientations of the simplex and its faces. For simplicity we represent all of these with the same symbol.}

Consider now a simplicial manifold $\Delta$ with boundary $\partial\Delta$. Its kinematical boundary Hilbert space is again obtained as the tensor product $\mathcal{H}_{\partial\Delta}^{\text{kin}} = \otimes_f \mathcal{H}_f$ over all the $(d-1)$-simplicial faces $f$ in $\partial\Delta$. The probability amplitudes for the boundary basis states $|{\bf n}_1 \ldots {\bf n}_{|\partial\Delta|}\rangle$ are then obtained by contracting the simplicial amplitude tensors $A_{{\bf n}_1 \ldots {\bf n}_{d+1}}$, one for each individual $d$-simplex in $\Delta$, with gluing tensors $G_{{\bf n}{\bf n}'}$ according to the connectivity structure of $\Delta$. These uniquely define the amplitude map $A_{\Delta}: \mathcal{H}_{\partial\Delta}^{\text{kin}} \rightarrow \mathbb{C}$. The physical boundary Hilbert space is again obtained as $\mathcal{H}_{\partial\Delta}^{\text{phys}} = \overline{\mathcal{H}_{\partial\Delta}^{\text{kin}} / \text{ker}(A_{\Delta}) }$.

\subsection{State-sum formulation for quantum gravity $\Rightarrow$ QFT vector models}
In QG, we want to perform a sum over all bulk manifolds which are compatible with a given boundary data. Let us therefore specify a closed $(d-1)$-dimensional boundary simplicial manifold $B$ with a boundary state $|\chi_B\rangle \in \mathcal{H}_{B}^{\text{kin}} = \otimes_f \mathcal{H}_f$, where the tensor product runs over all the simplicial faces in $B$. Then the QG amplitude for this boundary data is $A_{QG}(|\chi_B\rangle) = \sum_\Delta A_\Delta(|\chi_B\rangle)$, where the sum runs over all the simplicial manifolds $\Delta$ such that $\partial\Delta = B$. The sum over all simplicial manifolds can now be implemented as a `state-sum model' similarly to random tensor models \cite{Gurau2024} and group field theory \cite{Oriti09}.

Consider a random vector model type of partition function $Z(\lambda) = \int \left[\prod_{i=s,t} \text{d}\Psi^{(i)}\right] e^{-S[\Psi^{(i)};\lambda]}$, where $S[\Psi^{(i)}_{\bf n};\lambda] = \frac{1}{2}\sum_{i=s,t} G^{-1}_{{\bf n}_1 {\bf n}_2} \Psi^{(i)}_{{\bf n}_1} \Psi^{(i)}_{{\bf n}_2} + \lambda A_{{\bf n}_1 \ldots {\bf n}_{d+1}} \Psi^{(i_1)}_{{\bf n}_1} \cdots \Psi^{(i_{d+1})}_{{\bf n}_{d+1}}$,
and we sum over repeated Fock space indices. Here, $\Psi^{(i)}_{{\bf n}}$ ($i=s,t$) is a QFT state vector on a spacelike ($s$) or timelike ($t$) simplicial face. These are the dynamical variables of this statistical model --- hence the name `QFT vector model'. $\lambda$ is the coupling constant of the model. If we perform a perturbative expansion of $Z(\lambda)$ in the coupling constant $\lambda$, we will see that the Feynman diagrams that arise correspond exactly to different closed simplicial manifolds (including all possible topologies), which are obtained by gluing simplices together along their faces. Moreover, the Feynman amplitudes are exactly the amplitudes that we get by contracting the amplitude tensors according to the structure of the manifold with the gluing tensors, except that they are multiplied by a factor of $\lambda^{N_\Delta}/\Sigma_\Delta$, where $N_\Delta$ is the number of simplices in the simplicial manifold $\Delta$ and $\Sigma_\Delta$ is a symmetry factor --- an important point for later! Now, let $B$ again be a closed $(d-1)$-dimensional simplicial manifold with $N$ simplices and $|\chi_B\rangle\in \mathcal{H}_{B}^{\text{kin}}$ with vector components $\chi_{{\bf n}_1 \ldots {\bf n}_N}$. Then, the quantum gravity amplitude $A_{QG}(|\chi_B\rangle)$ is obtained as the expectation value $\langle \left[ \prod_{k=1}^N G_{{\bf m}_k {\bf n}_k} \Psi^{(i_k)}_{{\bf m}_k} \right] \chi_{{\bf n}_1 \ldots {\bf n}_N} \rangle_{\lambda\rightarrow 1}$ with respect to the partition function $Z(\lambda)$.

\section{Basic properties}
\subsection{Induced gravity amplitudes}
In \cite{raasakka25} we showed for the free scalar field theory in 2d simplicial Lorentzian spacetime that the gravitational Regge action arises approximately from QFT amplitudes as an effective low-energy (IR) action. We refer the reader to \cite{raasakka25} for details. Here we will have to settle with pointing out some general properties of the QFT model, which can give rise to approximate effective gravitational amplitudes in the IR.

First, let us assume that the QFT model is approximately free in the high energy (UV) limit. In this case, when $l_\Delta$ is small enough, we can approximate $G_n$ by a sum of products of $G_2$ via Wick's theorem. Correspondingly, the simplicial amplitude tensor $A_{{\bf n}_1 \ldots {\bf n}_{d+1}}$ from Eq.~(\ref{eq:amptens}) can be decomposed into a sum of products of amplitudes $B^{(ij)}_{kl}$ for processes having only two boundary excitations in modes $k$ and $l$ on faces $i$ and $j$ of the simplex, respectively. $B^{(ij)}_{kl}$ can also be interpreted as the amplitude for one excitation to propagate from face $i$ to face $j$, or vice versa. Choosing a UV cutoff scale $\Lambda$, we obtain the IR effective model by summing over the contributions to the full amplitude from processes involving length scales below $\Lambda^{-1}$. If we set $l_\Delta \sim \Lambda^{-1}$, then all excitations on a single face of the triangulation are above the cutoff scale except for the zero (constant) mode. The leading contribution to the amplitude associated with bulk $(d-2)$-simplices of a simplicial manifold are given by loops of single excitations propagating around the $(d-2)$-simplices. The multiplicative contribution to the full amplitude by such processes are of the form $\text{tr}((\tilde{G}\tilde{B})^n)$, where $n$ is the number of $(d-1)$-simplices sharing the $(d-2)$-simplex, and $\tilde G, \tilde B$ are the gluing and amplitude tensors for two boundary excitations with the zero mode removed. Now, if the matrix $\tilde{G}\tilde{B}$ has a dominant eigenvalue $x = |x|e^{i\varphi}$, then $\text{tr}((\tilde{G}\tilde{B})^n) \sim e^{i\varphi n}$. This will produce an effective gravitational (Regge) amplitude, since the deficit angle around a $(d-2)$-simplex is linear in $n$.

\subsection{Renormalization of the cosmological constant}
As we already mentioned, the main issue with the original induced gravity proposal was that the effective cosmological constant comes out way too large. The same issue can be observed for QFT on a simplicial manifold, as was shown in \cite{raasakka25}. However, in the QFT vector model formulation the amplitude for a manifold $\Delta$ gets multiplied by the factor $\lambda^{N_\Delta}/\Sigma_\Delta$, where $N_\Delta$ is the number of simplices in $\Delta$. Thus, any contribution $S' = c N_\Delta$ to the effective IR action proportional to the total volume of the manifold, and thus $N_\Delta$, can be absorbed into a redefinition of the QFT vector model coupling constant $\lambda \mapsto \lambda e^{-ic}$. Requiring $\lambda$ to be real-valued and positive already automatically sets any term in the effective action proportional to the total volume of the manifold to zero. Accordingly, the naive very large value for the effective cosmological constant poses no issue in the QFT vector model. It is an open question, however, if we can somehow recover the observed finite-but-very-small value for $\Lambda$ by some other mechanism.

\subsection{Classical and continuum limits}
It is interesting also to consider how to recover classical gravity in this framework.\footnote{I thank W.~Wieland for bringing this issue into my attention during my talk at the GR24-Amaldi16 conference. } In the naive classical limit $\hbar\rightarrow 0$ all quantum corrections vanish and the effective gravitational action goes to zero as well. In order to recover classical gravity in the limit, we need to simultaneously with the classical limit $\hbar\rightarrow 0$ take also the continuum limit $l_\Delta \rightarrow 0$ in such a way that the linear (in curvature) term in the effective gravitational action remains finite. In $d$ spacetime dimensions the effective gravitational constant $G_\text{eff} \propto l_\Delta^{2-d}$, so the first order term in the effective gravitational action is proportional to $\hbar/l_\Delta^{2-d}$. Thus, if we keep this ratio constant in the double-scaling limit $\hbar,l_\Delta \rightarrow 0$, we recover the classical gravitational dynamics in the limit. (The effective cosmological constant diverges in this limit, but this is not an issue in the QFT vector model, since we already absorbed it into the coupling constant.) From this perspective, we must consider $l_\Delta$ as a fundamental length scale in the model, since it determines the strength of the effective gravitational forces and provides us with a fundamental length scale instead of the Planck length.

\section{Summary and outlook}
We have discussed some basic definitions and properties of QFT in simplicial spacetimes and QFT vector models. Clearly, a lot more work is ahead of us to make the framework fully rigorous, well-defined and a serious candidate for a theory of QG. For example, there does not exist (to our knowledge) a standard way of doing QFT in bounded spacetime regions, and different quantization procedures can lead to different results. More realistic models beside the scalar field case, and higher dimensions than 2d, remain to be explored in any depth. Nevertheless, we hope to have convinced the reader that QFT vector models are at least an interesting new approach to QG and a fertile ground for further research.

\section*{Acknowledgments}
I would like to thank the organizers of GR24-Amaldi16 conference for giving me the opportunity to present my work, and by extension everyone who attended my talk. I received several excellent questions from the audience after my talk, some of which I hope to have answered better in this short paper.

%\bibliography{iopart-num}

\providecommand{\newblock}{}

\end{document}